\documentclass[10 pt, conference]{IEEEtran}
\usepackage{epsfig,graphicx,color,amsmath,amsfonts,bm,balance,tikz,array,multirow,fancyhdr, blindtext,multirow, algorithm, algpseudocode,bbm}
\usepackage{url}
\usepackage{amsmath}
\usepackage{float}
\usepackage{todonotes}

\title{\LARGE \bf
Predicting Bit Error Rate from Meta Information using Random Forests
}

\author{Jianyuan Yu, Yue Xu, Hussein Metwaly Saad and R. Michael Buehrer
\thanks{$*$This work was supported by Wireless @ Virginia Tech}
\thanks{$^{*}$}
\thanks{$^{\dagger}$%
       } \\
Bradley Department of Electrical and Computer Engineering, Virginia Tech, Blacksburg, VA 24061 \\

\{jianyuan,xuyue24, husseinm19,buehrer\}@vt.edu
}

\begin{document}

\maketitle
\thispagestyle{empty}
\pagestyle{empty}

\begin{abstract}
With the increasing power of machine learning-based reasoning, the use of meta-information (e.g.,  digital signal modulation parameters, channel conditions, etc.) to predict the performance of various signal processing techniques has become feasible. One such problem of practical interest is choosing a proper interference mitigation method based on the meta information of the received signal. Since heuristic table-based methods suffer from limited prediction capability for unseen cases, we propose a recommendation system based on the use of Random Forests (RF).  Specifically, RF used to predict the Bit-Error-Rate (BER) of all mitigation approaches so as to determine the approach with the best performance. We found RF can predict BER with high accuracy, and its importance factor demonstrates which input attributes matter most. These BER prediction results can also benefit other functions such as adaptive modulation, channel sensing, beaming selection, etc.
\end{abstract}


\section{INTRODUCTION}
With the exponential increase in spectrum usage, interference mitigation becomes essential. In the literature, several techniques and algorithms have been proposed to cancel or suppress various types of interference. For example, a filter bank approach ~\cite{chen2013non} utilizes a framework of non-maximally decimated filter banks with the property of perfect reconstruction which can substantially mitigate most narrowband interference. The notch filter   ~\cite{soderstrand1997suppression} creates a notch-like bandstop filter to efficiently remove narrowband interference. Additionally, the fractional Fourier transform (FRFT) can transform chirp interference into a peak in the fractional frequency domain which can be identified and eliminated ~\cite{dorsch1994chirp}. If spread spectrum is used by the signal of interest (SoI), then the demodulation/correlation process  can mitigate uncorrelated interference due to correlation properties of the spreading code ~\cite{wang2010narrowband}. Other methods including the constant modulus algorithm  ~\cite{Veen96} or a frequency shift filter ~\cite{adlard2000frequency} can also be applied to the problem. However, none of these techniques can mitigate all types of interference or do so equally well. Thus, a mitigation recommendation system which can quickly determine which technique  to apply based on a small number of simple features, i.e., \textit{meta information} can benefit wireless systems.  Potential meta information includes SoI modulation type and the number of bits per symbol,  as well as interference parameters such as  interference type, center frequency, bandwidth, etc. In addition to using the  predicted BER for determining the mitigation approach, it can also be forwarded to other applications such as channel hopping, beam selection, adaptive modulation etc. ~\cite{kaddoum2009generalized} ~\cite{nasser2008bit}.

The performance of the interference mitigation techniques  mentioned above heavily depends on the interference type.  Thus, the required meta information will depend on  the primary feature of interference type. For instance, the center frequency is critical for tone interference, while the chirp rate is a primary parameter for a chirp interference. Our previous work ~\cite{yu2020interference} proposed an interference classifier using spectrum information and achieved high accuracy. The work ~\cite{o2017introduction} summarizes the literature concerning  classifying or learning the modulation parameters using neural networks in a cognitive radio system. Further, there have been several recent works concerning adaptive recommendation engines for use in  cognitive radio ~\cite{volos2010cognitive} ~\cite{asadi2014learning}.  For example,  ~\cite{asadi2015metacognitive} uses meta information for reasoning and prediction, while ~\cite{asadi2014learning} adapts the $k$-nearest neighbor approach,  ~\cite{asadi2015metacognitive} uses belief propagation, and ~\cite{park2019learning} shows that meta information such as pilot information is sufficient to learn channel conditions in an IoT scenario. As to the specific task of predicting BER, previous work predicts the BER ~\cite{kaddoum2009generalized}  ~\cite{nasser2008bit} statistically  by deriving a  closed-form mathematical solution, but assumes that a large number of parameters are known {\em apriori}.  Further,  it is only applicable to typical OFDM modulated signals and interference, but is not applicable to other cases. 



In this paper, we propose a BER predictor for cognitive radio systems to determine the best narrowband interference mitigation technique. The proposed method  uses normalized meta-information as features, and accurately predicts the BER of all mitigation methods examined. Meanwhile, RF also offers an important factor to quantitatively measure which features matter the most, and we explore the relationship between prediction error with the most  important features. We show that RF is significantly better than a  heuristic table-based method, and we present a comparison between random forest to XGB, neural networks and Support Vector Machines (SVM). The simulation results show that our  predictor is promising for use in cognitive radio systems and can be easily extended to include in other mitigation methods or predict other metrics.


\section{System Model}

\begin{figure*}[h]
  \includegraphics[width=\textwidth,height=6cm]{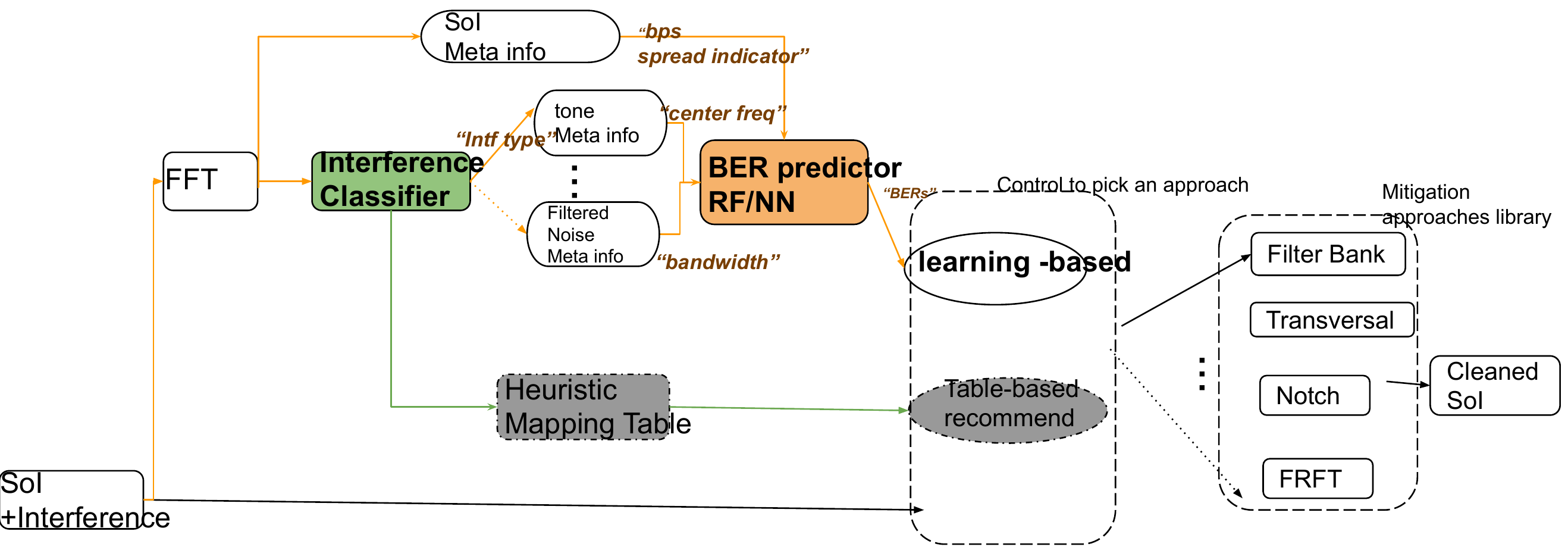}
  \caption{Flowchart of Interference Mitigation with Mitigation Recommender System}
  \label{fig:flowChart}
\end{figure*}

In this work, our goal is to design a smart controller to choose the best interference mitigation approach, as well as predict the resulting BER. The estimated BER could also be used for channel quality estimation, beam selection, or MIMO adaptation. Moreover, the BER can also serve as a warning indicator when all the mitigation techniques fail and the predicted BER is higher than a certain threshold.  Fig.~\ref{fig:flowChart} illustrates how the mitigation recommender system could be applied in a cognitive radio system employing interference mitigation. The received signal first is pre-processed by Fourier Transformation (FFT) into spectrum domain.  The pre-processed samples are then passed to two parallel paths for determining the  Signal-of-interest (SoI) meta-information and the interference meta-information.   The SoI block estimates information such as modulation, the type of SoI  (e.g., is DSSS modulation used or not) and the modulation order. The bottom path first passes the signal through an interference classifier which can classify  the type of  interference.  The signal is then passed through blocks to determine the interference meta information. The meta-information for both the SoI and interference is then  passed to the BER preditor, which outputs BERs of all interference mitigation techniques (including no mitigation).  Finally a  recommendation is provided by choosing the approach with the lowest predicted BER. As a baseline for comparison, we also include a table-based recommendation (shown in the grey boxes), which only requires the interference type and does not predict the exact BER value, and we will explain later how this table is formed.

\subsection{Signal and Interference Model}

The mathematical model of the continuous-time received signal  $r(t)$, is $ r(t) = x(t) + \frac{1}{\sqrt{2{J2S}} }i(t) +  \frac{1}{\sqrt{2{SNR}} } n(t),$
where $x(t)$ is the transmitted signal (SoI), $i(t)$ is the interfering signal, $n(t)$  additive-white Gaussian noise (AWGN), J2S is the jammer-to-signal power ratio, and SNR is the signal-to-noise ratio. The continuous-time received signal could be either standard single-carrier modulation  or  Direct-Sequence Spread Spectrum (DSSS).

Meanwhile, we consider five  types of interference $i(t)$, which are (note that the noise $n(t)$  is assumed to always be present):     1) {\bf None}: only the AWGN  $n(t)$; 2) {\bf Single Tone}: $i(t)$ is a complex sinusoidal signal with a constant frequency; 3) {\bf Chirp}: $i(t)$ is a complex sinusoidal signal with a frequency that is either linearly or exponentially changing with time.  The examples are generated by varying the frequency rate of change, known as chirpiness, which is defined as $\gamma = df / dt$. Therefore, each distinct value of $\gamma$ corresponds to a distinct spectrum; 4) {\bf Filtered noise}: $i(t)$ is similar to $n(t)$ but is passed through a low-pass filter given by $ H(z) = \frac{1 + (a-1)z^{-1} } {1+az^{-1}}$, where $a \in (0,1)$ is a parameter that controls the filter width, the larger the value of $a$, the wider the bandwidth; 5) {\bf Unknown modulated signal}: $i(t)$ is a randomly modulated, and information-carrying signal, while the modulation parameters are unknown, and its bandwidth is assumed to be much smaller than SoI. 

\subsection{Interference Mitigation Approaches}

\begin{figure*}[h]
 \includegraphics[width=\textwidth,height=8cm]{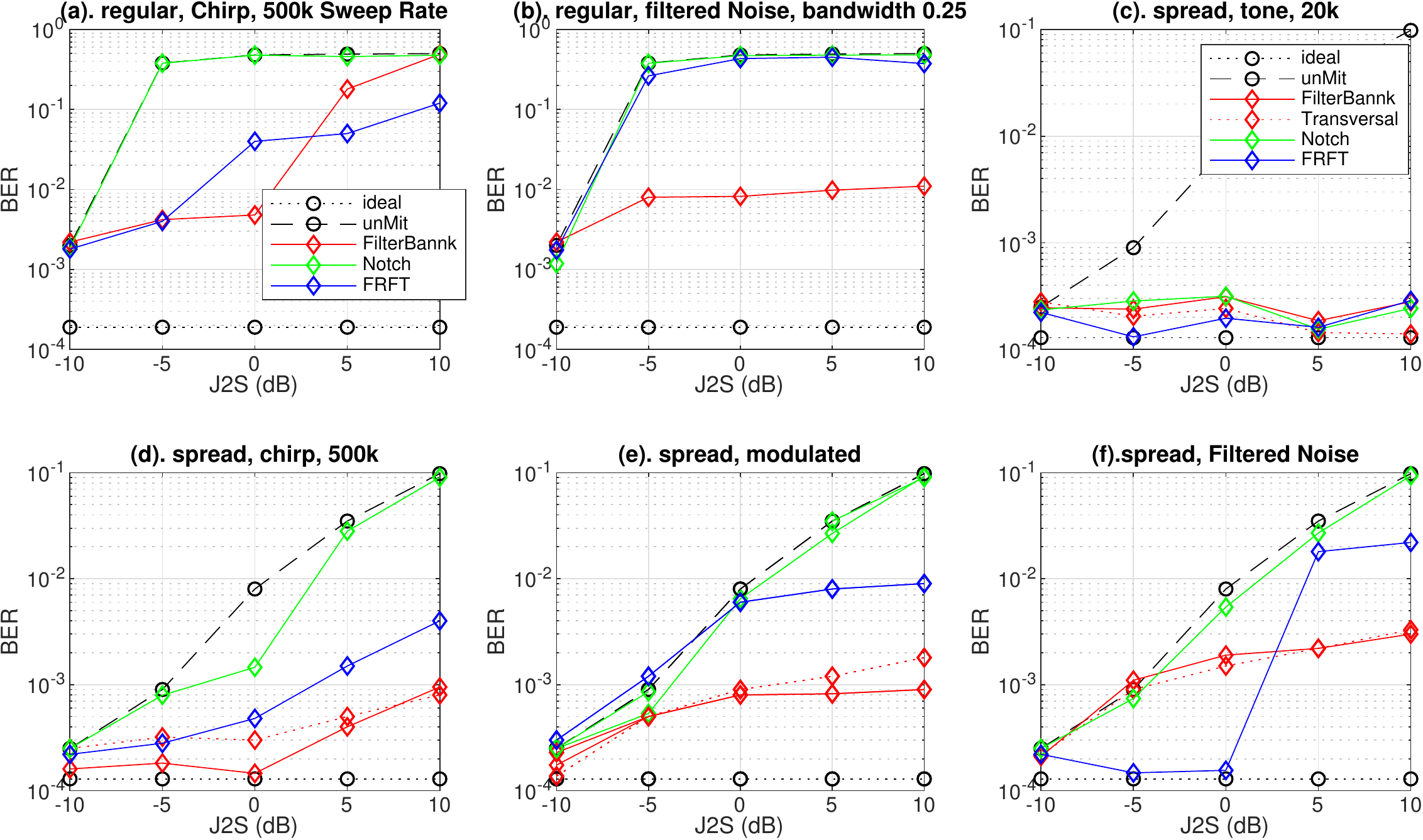}
  \caption{Simulated BER of Different Interference Mitigation Approaches for Different Interference Types}
  \label{fig:bers}
\end{figure*}

Four interference mitigation methods are implemented and briefly described here. Each has its own advantages and disadvantages mainly depending on the interference type or the use of spread spectrum.

\subsubsection{Filter Bank}
A filter bank ~\cite{chen2013non} is the most efficient method in our library, as it works for almost all types of interference. However, as we will describe later, there are some cases where the filter bank cannot provide satisfactory results, or requires more computation time and is thus less preferable.

\subsubsection{Transversal Filter (TF)}
It is notable that Transversal Filter only works for a DSSS SoI, and for all types of interference, Transversal Filter has similar results with Filter Bank but requires less computation ~\cite{wang2010narrowband}.

\subsubsection{Notch Filter}
Although this technique applies classic notch filtering, it also uses an adaptive technique ~\cite{soderstrand1997suppression} to determine the filter coefficients based on an {\em a priori} unknown tone frequency. Specifically, we implement a notch filter with the following transfer function
\begin{equation}
    H(z) = \frac{2-k_2}{2} \frac{1- \frac{2(2-k_2-k_1^2)}{2-k_2}z^{-1} + z^{-2}}{1-(2-k_2-k_1^2)z^{-1} + (1-k_2)z^{-2}},
\end{equation}
where $k_1$ and $k_2$ are related to the pole radius $r$ and digital notch frequency $\theta$, as $k_1 = \sqrt{1+r^2-2r\cos{\theta}}$ and $k_2 = 1-r^2$. The function requires the designer to specify the pole radius $r$ ̧ while the function adaptively estimates the appropriate notch frequency $\theta$ and the coefficients. The notch filter is easy to implement, as it requires the least computation. However, we notice that it is effective only for extremely narrowband  tone-like interference, and may not work well to suppress interference like filtered noise with a larger bandwidth.

\subsubsection{Fractional Fourier Transformation (FRFT)}
The discrete FRFT can be written as a matrix multiplication
\begin{equation}
    X_{\alpha} = F_{\alpha}x,
\end{equation}
\begin{equation}
    F_{\alpha}[m,n] = \sum_{k=1,k\neq N/2-1}^{N}{u_k[m] e^{-j\frac{\pi}{2}ka} v_k[n]},
\end{equation}
where $u_k[m]$ and $v_k[n]$ are the elements of the eigen matrix ~\cite{dorsch1994chirp}. The approach ~\cite{dorsch1994chirp} performs a search over-rotation angles by repeated calls to the fraction FT function, looking for a rotation that results in a large peak in the rotated time/frequency domain. By nulling the signal at that location and rotating the signal back, we can eliminate chirp signals. The FRFT approach can suppress tone and chirp interference with both narrowband and DSSS SoIs. However, the FRFT needs to exhaustively search for the correct rotation  $\alpha$ to convert the chirp to "tone" in the fractional spectrum domain, and the search includes a minimum step size when choosing  $\alpha$. Depending on the chirp rate, it may have to take a small step size and thus many iterations to find the optimal value of $\alpha$.

Fig.~\ref{fig:bers} shows simulated BER results for the different mitigation approaches and interference types. In each plot, the black lines denote the ideal BER in the absence of  interference and the BER in the absence of any mitigation (i.e., lower and upper bounds). In plot Fig.~\ref{fig:bers}(a), we see that the filter bank can mitigate chirp interference at a low chirp rate, while the FRFT approach works better at a higher chirp rate and J2S. In plot Fig.~\ref{fig:bers}(b), we see that all of the  methods are effective with DSSS SoI and  tone interference, although the notch filter works the fastest and hence most preferable. In plots Fig.~\ref{fig:bers}(c) to (f), we see that both the FRFT and TF work well, but TF requires much less computation and is thus preferable. We define a technique to be ``effective'' if its resulting BER is under 1\%.  Thus, there are few cases where no technique is effective, such as the case of high-chip-rate chirp interference and high J2S in Fig.~\ref{fig:bers}.a. In this case, the high BER can be passed to the cognitive radio to change other parameters (e.g., frequency band) to ensure acceptable  performance.

\begin{table}[h]
\caption{Heuristic table mapping interference to best approach}

\label{table:map}
\begin{center}
\begin{tabular}{|c|c|c|c|c|}
\hline
\footnotesize{} & tone & chirp & modulated & filtered Noise   \\
\hline
regular  & Notch & FRFT & filter Bank  & filter Bank \\
\hline
spread & Notch & Filter Bank & Transversal  & Transversal  \\
\hline
\end{tabular}
\end{center}
\end{table}

\subsection{Heuristic Mapping Table Approach}
Based on the previous plots and observation we can derive a heuristic mapping table in Table.~\ref{table:map}, as to pick a mitigation approach only by interference type and SoI type. This approach is useful most of the time, with notable exceptions, e.g using the filter bank approach at high J2S in the presence of chirp interference. There may be other cases not illustrated above, and it can be difficult to create a mapping table which will work for all cases. Moreover, unlike the RF predictor, it does not provide BER before running the mitigation method, and hence it cannot provide a warning indicator before applying a  mitigation technique.

\subsection{ Data Preprocessing }
Most supervised learning approaches use features that have similar types. However, our meta information is a mixture of data types, including both  discrete features such modulation type, as well as continuous features such as center frequency. Another challenge is that the BER is unevenly distributed, meaning that  in certain ranges there is limited training data available. To this end, we explore several aspects here: 
    \subsubsection{Training Data} Most of our mitigation methods work in the frequency domain, which means the raw data or its Fourier transformation contains important information. However, each example of the raw data is very large (up to $10^6$), which makes it  cumbersome for training purposes. As a contrast, our results show that the lightweight meta-information is sufficient to predict the BER well with much less training size.
    \subsubsection{Normalization} The parameters of each input vary substantially, e.g modulation order is either 1 or 2, while the sweep rate of a chirp ranges from 1e5 to 500e5 Hz/s. We first normalize these by the sample rate, then normalize them into range $[0,1]$ via $\textbf{x}' = \frac{\textbf{x}-min(\textbf{x})}{max(\textbf{x})}$. 
    \subsubsection{Integer-to-Binary conversion} Some features are integers, e.g., bit-per-symbol (bps) of modulation order, but still have some meaning when the input is a floating point number. For example, we can obtain $BER_1$ when bps is 1 and $BER_2$ when bps is 2. When the bps is set as 1.5 for some reason, the predictor can still output $BER_3$ which will typically fall between $BER_1$ and $BER_2$. However, this is not true for category features, e.g., the interference type. Thus, if we assign a value of 2 for tone interference and 3 for  chirp interference, a value 1.5 does not carry any meaning.  To deal with this, we propose to convert integers to a vector of binary values. In our case, the interfering type can be described by a 3-digit binary vector.  
    \subsubsection{Skewed data distribution} Our previous analysis of these methods shows their BER distribution varies significantly and hence there are  some regions where few data points exist. For example there are few points in the range [-10,0]dB. For filter bank in its histogram, since it usually works well and the BER is pretty low. Our later result shows the skewed dataset is not a significant problem for the random forest and can be ignored. 
    \subsubsection{Missing data} The last issue is  missing or invalid data, which comes from two cases: (a). The N/A case of TF mitigation for narrowband signals. (b). The blank fields in  interference parameters when only some of them are applicable. To handle the former case, we used a BER 0.50. For the latter case, the missing interference parameters were filled with a default value, which is the most frequent value in the current dataset. Both approaches show to work well in  our simulations.



\subsection{Random Forests (RF)}

\begin{table}[tbpt]
\caption{Computation Complexity}
\label{table:complex}
\begin{center}
\begin{tabular}{|c|c|c|}
\hline
Methods & train & test\\
\hline
NN & $ O(N N_{it} \sum_{i=1}^{K}{ L_i L_{i-1}} )$ & $O(N \sum_{i=1}^{K}{ L_i L_{i-1}} )$ \\
\hline
RF & $O(MPN\log(N))$  & $O(MP)$\\
\hline
\end{tabular}
\end{center}
\end{table}
An important branch of supervised learning is decision tree-based techniques, including random forests and extreme gradient boost tree. The concept of a decision tree is that it predicts using a tree-like structure, where a leaf denotes a value of features and the branch represents the weight, and hence decision trees represent a disjunction of
conjunctions of constraints on the attribute values of
instances. The objective is to minimize the training loss $l$ and model $\Omega$, 
\begin{equation}
    \min{ ( \sum_{i=1}^{N} l(y_i,\hat{y_i}) +  \sum_{k=1}^{P} \Omega (x_k) )},
\end{equation}
where $N$ is the number of datasets, and $P$ is the number of features. The first term $l$ measures how well the model fits the training data and the second term $\Omega$ measures the complexity of trees. Single decision trees  suffer from overfitting and high variation,  random forest have been proposed to ease the variation by referring to multiple random trees and partially choosing  features to suppress the correlation problem ~\cite{breiman2001random}. The random forest chooses the most voted result among multiple decision trees. Alternatively, rather than generating decision trees in a random way, the gradient boosted tree gradually generates a new tree based on  previous trees and the loss function.  Further, XGB  speeds up the training by calculating the optimization problem in the form of  Talyor series. The complexity of RF  ~\cite{liaw2002classification} as compared to NNs is shown in Table.~\ref{table:complex}. As can be seen, RF requires significantly less computation both in terms of training and testing. For NN, $N$ is the train/test size, $K$ for a number of layers, $L_i$ is the number of neurons in the $i$-th layer. For RF, $M$ is the number of trees (we use a value 200), $P$ is the number of features, which is 14 in our work. 

The training includes the placement of features on the  nodes in each tree, which is related to the importance of the features. The more important features tend to be closer to the root and added into the tree earlier. The importance of a variable $X_j$  ~\cite{breiman2001random}, or mean decrease of impurity (MDI), for an ensemble of $M$ trees $\phi_m$ is measured by 
\begin{equation}
    MDI(X_j) = \frac{1}{M} \sum_{m=1}^{M} \sum_{ t\in \phi_m}{ \mathbbm{1}{(j_t=j)} [ p(t) \Delta i(t) ] }, 
\end{equation}
where  $p(t) = N_t/N$ denotes the variable used at node $t$, and
$\Delta i(t)$ is the impurity reduction at node $t$:
\begin{equation}
    \Delta i(t)  = i(t) - \frac{N_{t_L}}{N_t}i(t_L) - \frac{N_{t_R}}{N_t}i(t_R),
\end{equation}
where $N_t$ is the number of nodes under the subtree rooted at node $t$, while $N_{t_L}$ and $N_{t_R}$ is the left and right subtree size. As we will show later MDI can tell us the distribution of prediction error over certain features.

\section{Performance Evaluation}
 We used MATLAB to generate datasets, and Python  to build the neural network or random forest methods from \textit{scikit-learn} library. All of the dataset and source codes are publicly available in Github repository with detailed descriptions \footnote{https://github.com/yujianyuanhaha/BerPredict} and the results are reproducible.


\subsection{Dataset Generation}

As shown in Alg.~\ref{alg:sim} above, to generate the dataset, we iterate over J2S and SoI type, then in each loop, we iterate 100 times, where the inner loop includes randomly choosing an interference type, and generating signals contaminated with interference. Then the NN classifier determines which interference is present, and loads the corresponding tools to extract the interference parameters. Next, the RF predicts all the BERs and picks the lowest one, and the approach is applied. All the metadata and BERs are saved  for further evaluation. Our results are based on a dataset  size of 8000, with 20\%  used for testing, and the remaining for training. RF does not require validation data, while NN needs 20\% validation data out of the training set.  

The number of features is 14, and a  description and range of each feature value are shown in Table.~\ref{table:input}. We applied the method previously discussed: 1). Normalize the value by the min and max so it lies in the range [0,1], 2). Convert interference type from integer to binary, 3). Replace missing interference  parameters with the default value. Also, the BER of TF with a narrowband SoI is assigned a value of 0.5. The ideal BER is not included, since it is a constant  related only to SNR. The output is of size 5, including the BER without any mitigation applied (i.e., unmitigated), and 4 BERs corresponding to the mitigation approaches.

\begin{table}[tbpt]
\caption{Simulated Dataset Parameter Settings}
\label{table:input}
\begin{center}
\begin{tabular}{|c|c|c|}
\hline
ID & Setting & Range\\
\hline
1 & modulation rank & 1,2\\
\hline
2 & Jamming-to-Signal (J2S)  & (-10,10)(dB)\\
\hline
3 & Signal-to-Noise Ratio (SNR) & (8,12) (dB)\\
\hline
4 & interference type & 1 to 5\\
\hline
5 & duty cycle & [0.2, 1]\\
\hline
6 & single-tone center frequency & (1e3,20e3) Hz\\
\hline
7 & chirp sweep rate & (1e3,500e3) Hz/s\\
\hline
8 & unknown modulation, bps & 1,2\\
\hline
9 & unknown modulation, sps & (100,800)\\
\hline
10 & unknown modulation, bandwidth ratio & (0.025, 0.25)\\
\hline
11 & filtered noise bandwidth ratio & (0.8, 8e3)\\
\hline
12 & DSSS/spread indicator  & 0,1 \\
\hline
\end{tabular}
\end{center}
\end{table}



\subsection{BER Prediction}
\begin{figure}[h]
  \centering  \includegraphics[width=0.5\textwidth]{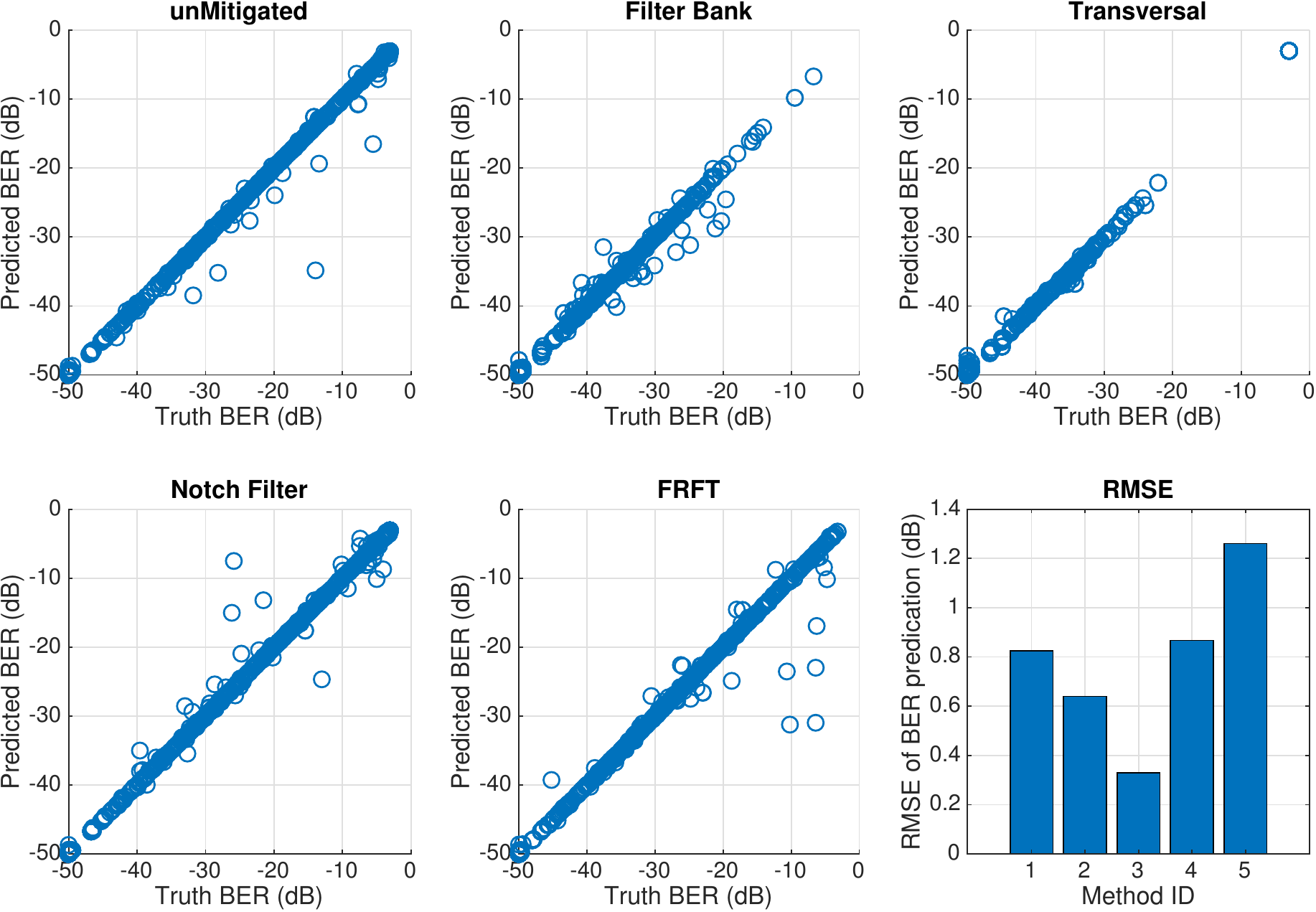}
  \caption{scatter plot of BER prediction for all approach}
  \label{fig:scatter}
\end{figure}

The performance of the RF predictor is shown in Fig.~\ref{fig:scatter}, where the five scatter plots show the predicted BER versus the ground truth (plotted as $10\log(BER)$ ), and we see there is nearly a $\hat{y}  = y$ line with a few outliers. The scatter plots also match the Root Mean Square Error (RMSE) histogram, where Filter Bank and TF usually perform well, and have fewer points at high BER. Here the RMSE is defined as $\sqrt{ \sum_{i=1}^{N}{ (y_i-\hat{y_i})}/N )}$ and $y$ is BER in logarithmic. The $6^{th}$ subplot shows the TF and Filter Bank also have better prediction than the other approaches. 

\subsection{Feature Importance}

\begin{figure}[h]
  \centering
  \includegraphics[width=0.30\textwidth]{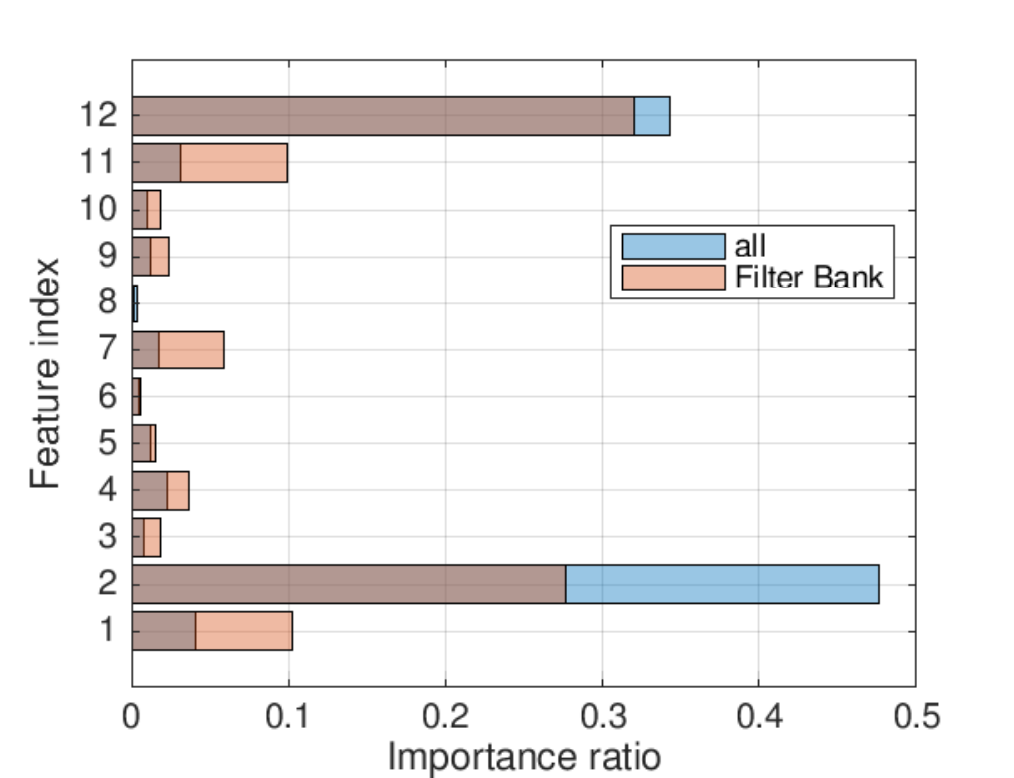}
  \caption{Random Forest Importance Ratio ( Index = parameters in Table 2)}
  \label{fig:FI}
\end{figure}

\begin{figure}[h]
  \centering
  \includegraphics[width=0.30\textwidth]{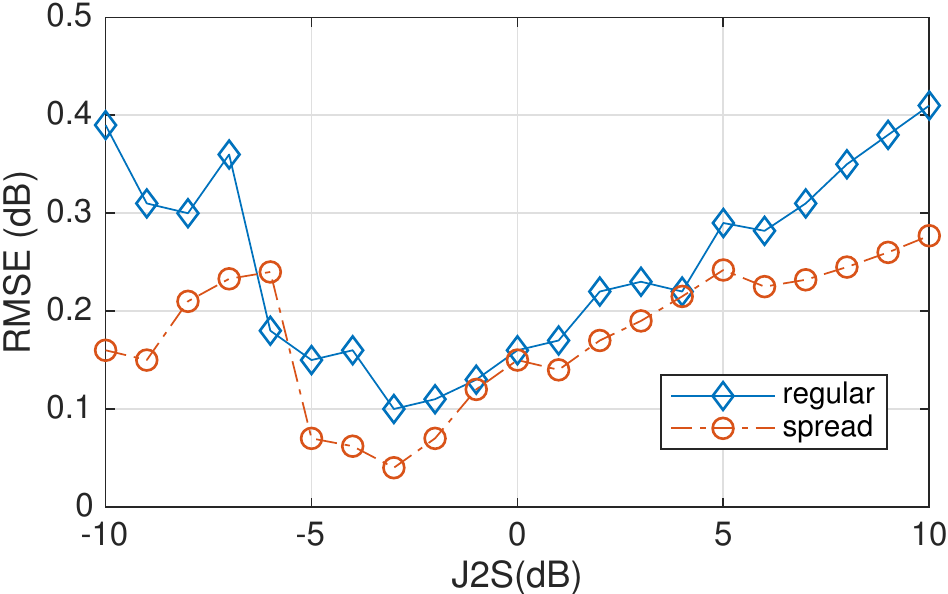}
  \caption{Estimation Error versus J2S for Both SoI Types}
  \label{fig:err}
\end{figure}

The RF method not only outputs the BER, but also gives an indicator of how important one feature is among all the input features. We can see in Fig.~\ref{fig:FI}, the 2nd feature (J2S) plays the biggest role  followed by the 12th  (DSSS SoI or not), while the others are much less significant. Here only the Filter Bank approach is plotted along with the combined.  The Filter Bank's  feature ranking is essentially the same as the combined results, but the values are more evenly distributed. 
In some scenarios, we are also interested in how estimation error is distributed. Based on the previous importance ranking, the RMSE versus J2S for both SoI types (DSSS or not) is shown in Fig.~\ref{fig:err}, we can see the error increases when the J2S is either too low or too high. Further, a spread SoI tends to have a lower estimation error than narrowband SoI. 

\subsection{Overall System Performance}
\begin{figure}[h]
  \centering
  \includegraphics[width=0.30\textwidth]{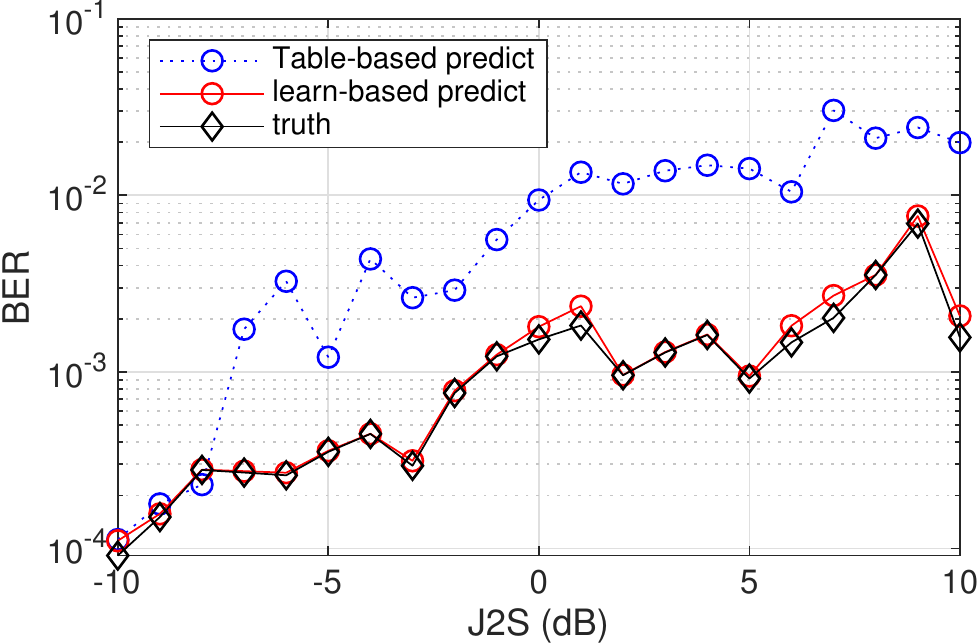}
  \caption{Performance of Table-based  and learning-based approach}
  \label{fig:ber}
\end{figure}

The overall performance versus J2S is shown in Fig.~\ref{fig:ber}. We can see our recommendation system matches the ideal system (labeled ``truth''), meaning that the predictor always provides the  approach with the lowest BER. The table-based approach results in significantly higher BER, especially at high J2S. 

\begin{table}[h]
\caption{Prediction RMSE of different predicting methods and mitigation approach}

\label{table:result}
\begin{center}
\begin{tabular}{|c|c|c|c|c|c|}
\hline
\footnotesize{} & unMit & FilterBank & TF & Notch  & FRFT \\
\hline
\textbf{RF}   & \textbf{0.12} & \textbf{0.16} & \textbf{0.10}  & \textbf{0.15}  & \textbf{0.18}\\
\hline
DNN  & 0.49 & 0.42 & 1.67  & 0.55 & 0.48 \\
\hline
XGB  & 0.62 & 0.55 & 1.92  & 0.55 & 0.69\\
\hline
SVM & 2.12 & 2.44 & 3.17  & 3.36  & 3.01\\
\hline
\end{tabular}
\end{center}
\end{table}

Besides the RF, we also examined the performance of several other methods like Deep Feed-Forward Neural Network (DNN), Extreme Gradient Boosting (XGB), and Support-Vector Machine (SVM) in Table.~\ref{table:result}. The DNN uses  $64\times 4$ hidden layers, the \textit{relu}  activation function, \textit{adam} as its optimizer and early stopping with patient count 5. The SVM uses \textit{radius} as a kernel function. RF with the number of trees equal to 200 provides the best results, while the DNN is better than XGB and SVM. The reason RF performs better than XGB is that RF trained fully grown decision trees, where XGB is based on weak learners that tend to result in shallow decision trees. Additionally, the distance function of SVM does not have physical meaning in our case which is a mixture of different data types, which explains why it performs the worst. We notice that while the BER of different mitigation techniques have different distributions, the RMSE does not differ much over the mitigations techniques.

\section{Conclusions and Further Work}
We have shown that using a Random Forest can accurately predict the BER using the dataset composed of meta-information such as modulation order and interference bandwidth. It shows that the mitigation methods' BER can be well estimated by meta-information using Random Forests. Moreover, the importance factor given by RF can tell us which input matters the most among all the meta information, as well as how estimation error distributes over these top factors. Such a recommender system could be applied to a cognitive radio to help recommend the best approach to apply to mitigation interference, and we show its performance is better than a  heuristic table-based one. Our future work includes expanding to include OFDM and MIMO waveforms, refining the cost function in training and including consideration of the computation cost of each approach, and exploring semi-learning when insufficient labels are available.

\addtolength{\textheight}{-12cm}  

\balance
\bibliographystyle{IEEEtran.bst}
\bibliography{IEEEabrv,references.bib}

\end{document}